\documentclass[prl,twocolumn,showpacs,preprintnumbers,aps,nofootinbib]{revtex4-1}
\usepackage{graphicx}
\usepackage{dcolumn}
\usepackage{bm}
\usepackage{amsmath,amssymb,amsfonts}
\usepackage{latexsym}
\usepackage{color}

\begin{document}


\title{\boldmath
Searching for new scalar bosons via triple-top signature in $cg \to tS^0 \to tt\bar t$
}

\author{Masaya Kohda$^1$, Tanmoy Modak$^1$, Wei-Shu Hou$^{1,2}$}
\affiliation{$^1$Department of Physics, National Taiwan University, Taipei 10617, Taiwan}
\affiliation{$^2$ARC CoEPP at the Terascale,
School of Physics, University of Melbourne,
Vic 3010, Australia}
\bigskip


\begin{abstract}
The alignment phenomenon, that the 125 GeV $h^0$ boson so
resembles the Standard Model Higgs boson, can be understood
in a two Higgs doublet model without discrete symmetry.
New Yukawa couplings $\rho_{tt}$ and $\rho_{tc}$ offer new avenues
to discover the extra scalar $H^0$ and pseudoscalar $A^0$.
We propose to search for $cg \to tH^0$, $tA^0$
followed by $H^0$, $A^0 \to t\bar t$, $t\bar c$,
where same-sign dileptons could be the harbinger, with triple-top,
in the signature of three leptons plus three $b$-jets, as confirmation.
Discovery could touch upon the origin of baryon asymmetry of the Universe.
\end{abstract}


\maketitle

\paragraph{Introduction.---}

With discovery of the 125 GeV boson $h^0$~\cite{h125_discovery},
existence of a second weak scalar doublet seems plausible,
especially since known fermions all come in three copies.
The $h^0$ boson resembles closely~\cite{Khachatryan:2016vau}
the Higgs boson of the Standard Model (SM).
For usual two Higgs doublet models (2HDM),
such as 2HDM~II that arises with supersymmetry,
this ``alignment'' phenomenon suggests that
the second doublet $\Phi'$ is rather heavy,
i.e. in the decoupling limit~\cite{Gunion:2002zf}.
While consistent with absence of beyond SM (BSM) physics
so far at the Large Hadron Collider (LHC),
having $m_{\Phi'}$ much above the TeV scale~\cite{Athron:2017yua}
is in strong contrast with $m_{h^0} \cong 125$ GeV.

It was pointed out~\cite{Hou:2017hiw} recently that,
if one drops the usual discrete $Z_2$ symmetry that forbids
flavor changing neutral Higgs (FCNH) couplings~\cite{Glashow:1976nt},
then alignment automatically emerges when
all extra Higgs quartic couplings are ${\cal O}(1)$:
extra Higgs bosons, sub-TeV in mass, could have
extra Yukawa couplings shielded from us by alignment.
Most interesting would be the diagonal $\rho_{tt}$
and the FCNH $\rho_{tc}$ couplings,
where $\rho_{tt}$ is the combination of two Yukawa couplings
orthogonal to the one that gives $y_t = \sqrt2 m_t/v \cong 1$,
with $v$ the vacuum expectation value (VEV).
Thus, $\rho_{tt} \sim {\cal O}(1)$ is plausible.

Since ${\cal O}(1)$ extra Higgs quartic couplings can produce
the needed first order electroweak phase transition (EWPT),
it was shown~\cite{Fuyuto:2017ewj} recently that
a complex $\rho_{tt}$ order 1 in strength can drive
electroweak baryogenesis (EWBG) rather efficiently.
In case $|\rho_{tt}|$ is accidentally small,
$\rho_{tc}$ could be a second option for EWBG,
if it is ${\cal O}(1)$ in strength and with near maximal phase.

Sizable $\rho_{tt}$ motivates one to consider
$gg \to S^0 \to t\bar t$ search, where $S^0 = H^0,\, A^0$
are the $CP$-even and odd scalars of 2HDM, respectively.
This search is, however, hampered by interference~\cite{Carena:2016npr}
with the large, underlying $gg \to t\bar t$ production, although
a recent ATLAS study~\cite{Aaboud:2017hnm} is indicative of experimental capabilities.
For $gg \to S^0 \to t\bar c$ (and $\bar tc$) search, Ref.~\cite{Altunkaynak:2015twa}
suggests that existing 13 TeV data can already probe down to $\rho_{tc} \lesssim 0.1$
for scalar mass below 400 GeV, and sensitive to larger $\rho_{tc}$ values at higher mass.
This search is akin to $s$-channel single top (i.e. $t\bar b$) search,
and ATLAS and CMS experiments are encouraged to update to 13 TeV.

\begin{figure}[b!]
\center
\includegraphics[width=.375\textwidth]{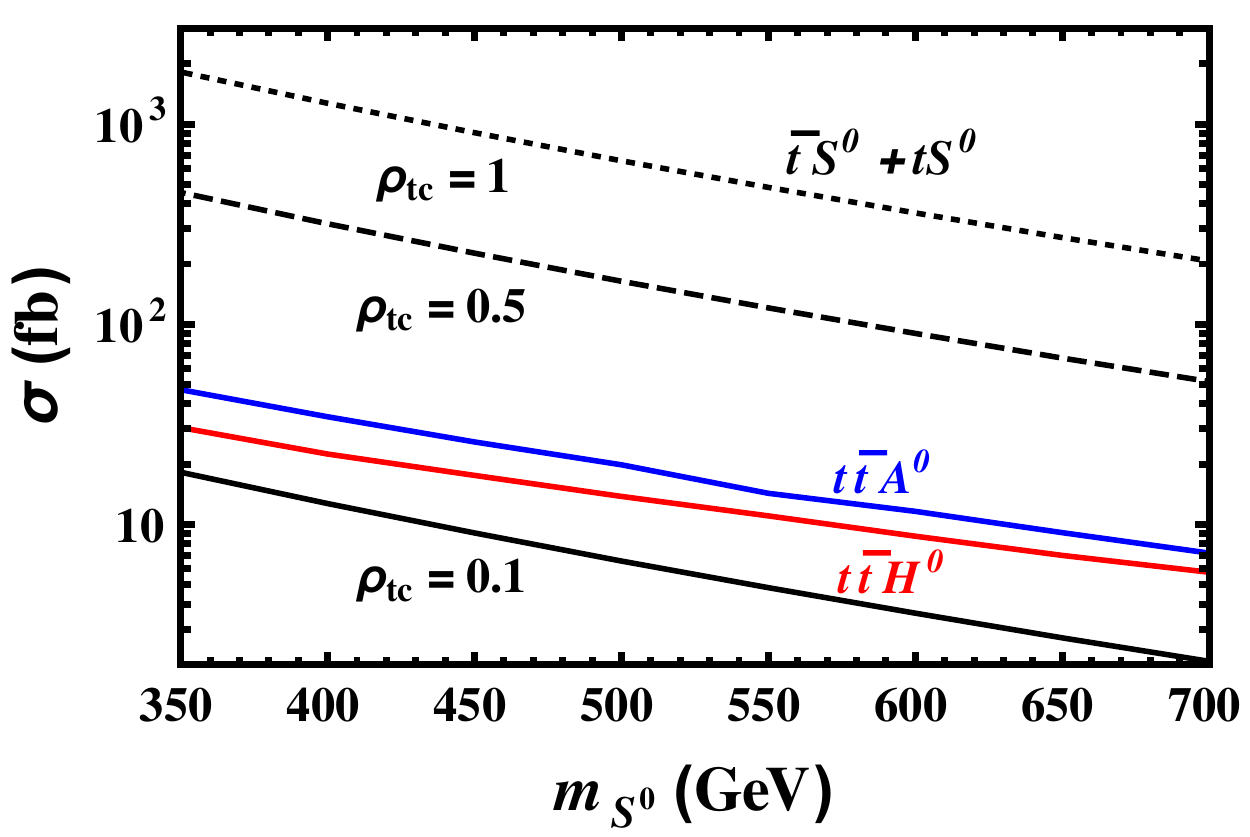}
\caption{
 Cross sections at $\sqrt s = 14$ TeV for
 $pp \to tS^0$, $\bar tS^0$ where $S^0 = H^0,\, A^0$,
 for $\rho_{tc} = 0.1$ (solid), 0.5 (dashed) and 1 (dots),
 and $pp \to t\bar tH^0$, $t\bar tA^0$ (for $\rho_{tt}=1$) as marked.
}
\label{partcross}
\end{figure}

%
\begin{figure*}[htbp!]
\center
\includegraphics[width=.325 \textwidth]{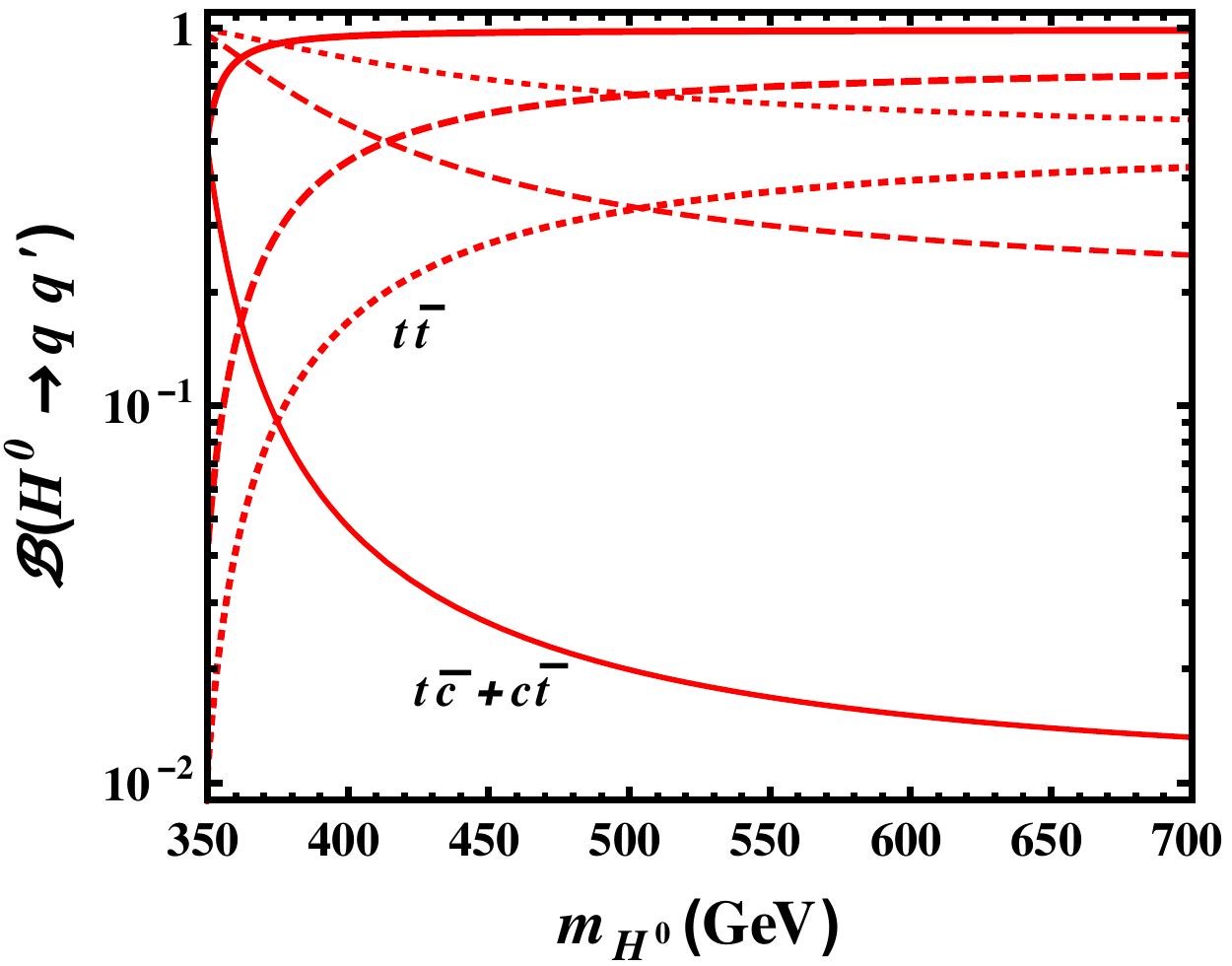}
\includegraphics[width=.325 \textwidth]{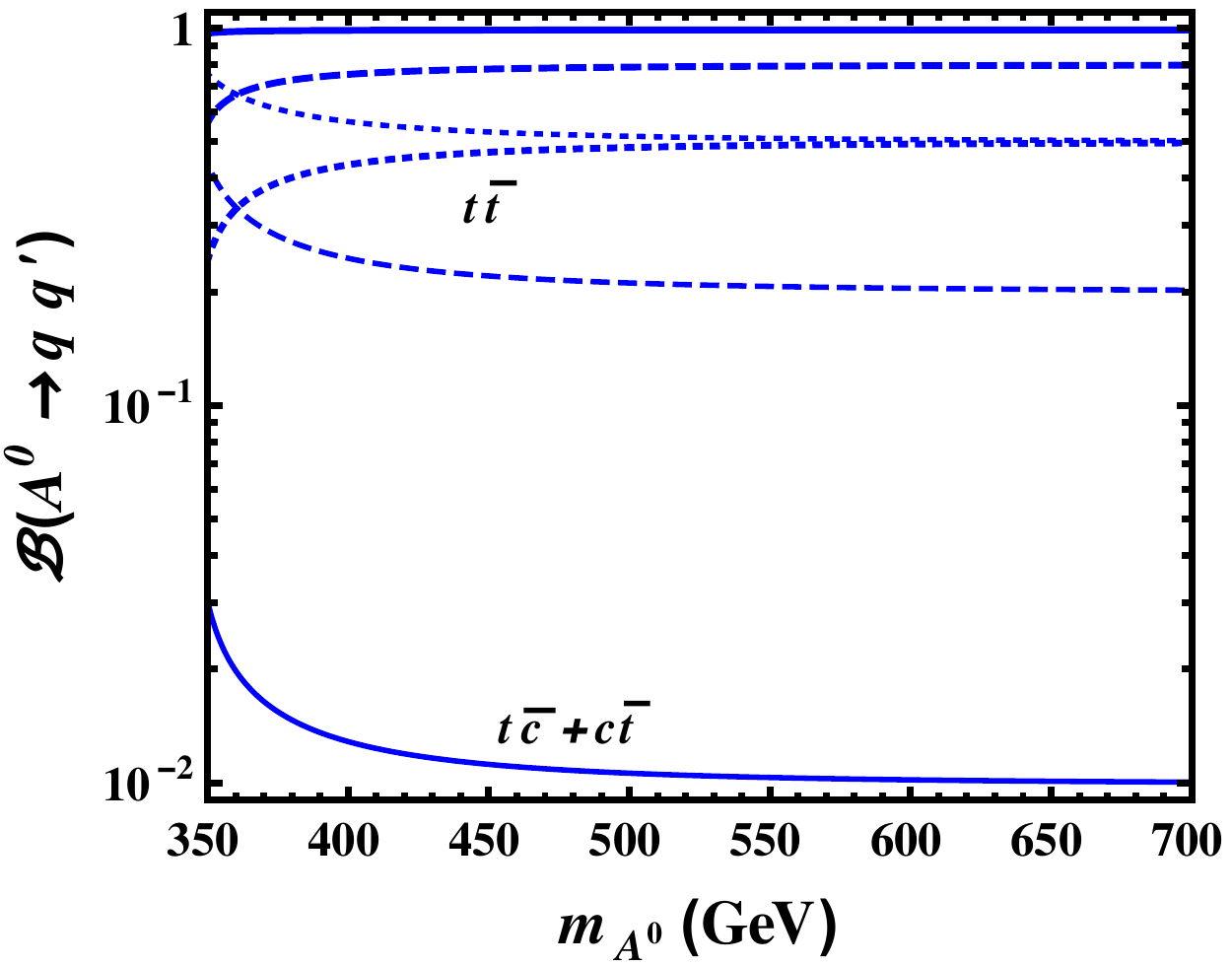}
\includegraphics[width=.318 \textwidth]{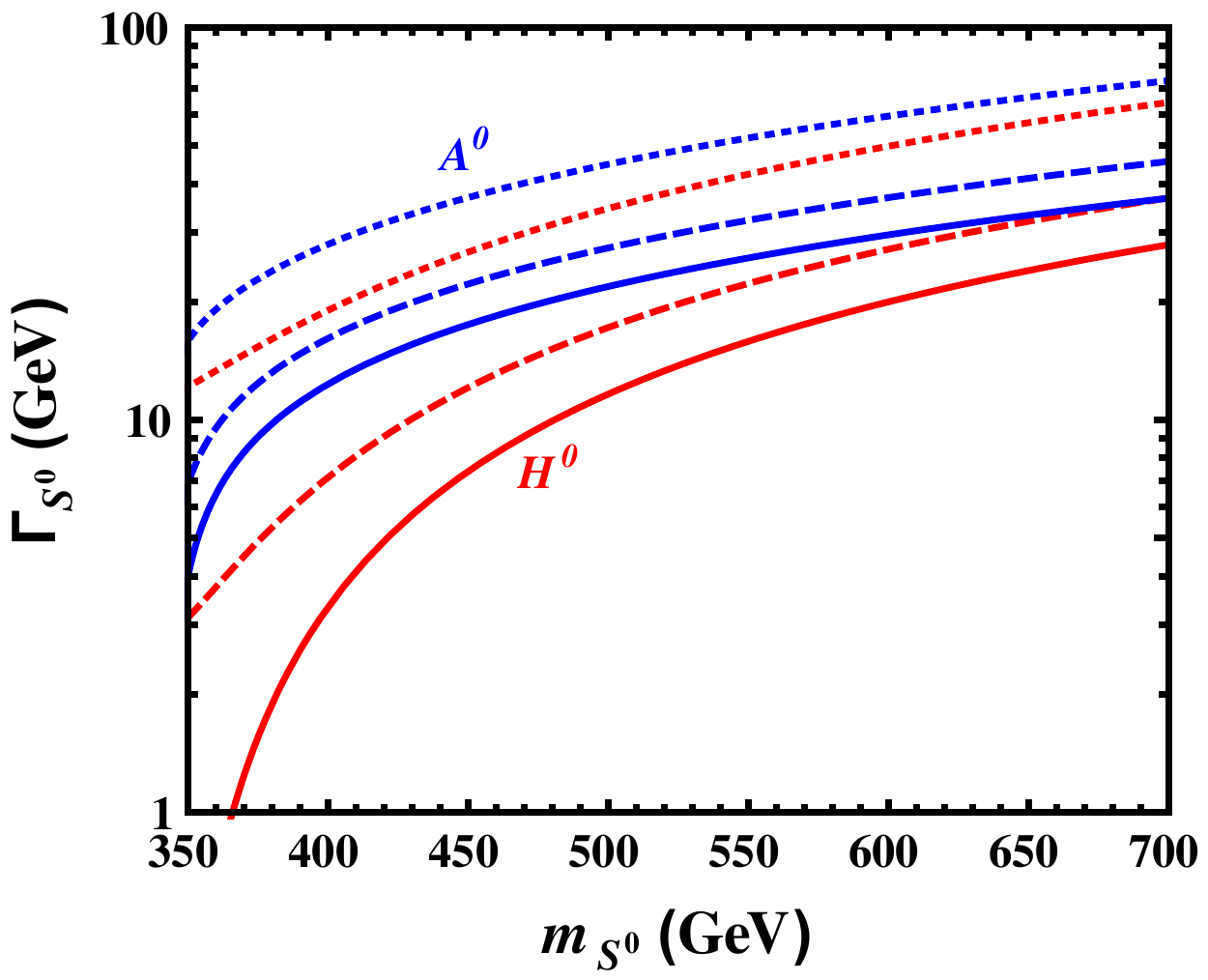}
\caption{
 Branching ratios for $H^0$ [left], $A^0$ [center]
 into $t\bar t$ and $t\bar c + \bar tc$, for $\rho_{tt} = 1$
 and $\rho_{tc} = 0.1$ (solid), 0.5 (dashed) and 1 (dots),
and corresponding widths [right], where $\Gamma_{A^0} > \Gamma_{H^0}$
for each $\rho_{tc}$.
}
\label{br-dw}
\end{figure*}
%

%
%

Even the $t\bar c$ resonance search, however, would
suffer from $t + j$ ($t\to b\ell\nu$) mass resolution,
which could approach~\cite{KFC, CMS:2017oas} 
200 GeV, depending on background.
As the cross section is considerably below $s$-channel single top,
the road to discovery could be long.
With emphasis on clear discovery, in this Letter
we suggest to search via $cg \to tS^0$~\cite{Iguro:2017ysu, Gori:2017tvg},
in particular in the triple-top or $tt\bar t$ final state.
This is because the SM cross section at fb order~\cite{Barger:2010uw}
is far below $\sigma(cg \to tS^0) \sim$ pb level for $\rho_{tc} \sim 1$,
which we display in Fig.~\ref{partcross} for $m_{S^0} \in (350,\, 700)$ GeV.

The \emph{tree level} $2 \to 2$ process $cg \to tS^0$ is
not too different in cross section from the loop generated
$gg \to S^0$ process, but the signature is more exquisite.
The precursor or harbinger may in fact come in same-sign dilepton plus jets (SS2$\ell$),
which we also study and find $cg \to tS^0 \to tt\bar c$ to be rather important.
But this signature could arise from a plethora of BSM physics.
The $3b3\ell$ signature, or ``triple-top'', would be less equivocal,
with $cg \to tS^0 \to tt\bar t$ a likely leading process.



\paragraph{Extra Yukawa interactions.---}

Following Refs.~\cite{Hou:2017hiw} and~\cite{Fuyuto:2017ewj},
we take the Higgs potential, which we do not display, to be $CP$-conserving.
The observed alignment, that $h^0$ resembles closely the SM Higgs boson,
means the mixing angle $\cos\gamma$
(analogous to $\cos(\beta - \alpha)$ in 2HDM~II)
between $CP$-even bosons $h^0$ and $H^0$ is rather small.
Taking $\cos\gamma \to 0$, the alignment limit,
$h^0$ turns into the SM Higgs boson that generates all masses,
while
\begin{align}
m_{H^0}^2,\, m_{A^0}^2 & = m_{H^\pm}^2 + \frac{1}{2}(\eta_4 \pm \eta_5)v^2, \\
m_{H^\pm}^2 & = \mu_{22}^2 + \frac{1}{2}\eta_3v^2,
\end{align}
where $\mu_{22}^2 > 0$ is the inertial mass of the heavy Higgs doublet $\Phi'$,
and $\eta_3$, $\eta_4$ and $\eta_5$ are Higgs quartic couplings (see Ref.~\cite{Hou:2017hiw}).
If the $\eta_i$s are all positive, which would facilitate alignment,
then $m_{H^0} > m_{A^0}$. But this condition is not necessary.
The principle raised in Ref.~\cite{Hou:2017hiw} is to have
all dimensionless parameters in Higgs potential of ${\cal O}(1)$, including $\mu_{22}^2/v^2$.
This is precisely the condition needed for EWBG~\cite{Fuyuto:2017ewj},
at least for generating first order EWPT.
Since $\mu_{22}^2$ is a decoupling mass,
not allowing it to damp away EWBG automatically implies sub-TeV $m_{H^0}$, $m_{A^0}$.
In fact, if one wishes to have the $\eta_i$s to be within some perturbative range,
then these extra scalars should not be much heavier than 500 GeV.

Having motivated the range for $m_{S^0}$, we shall take 
$\cos\gamma \to 0$ (and $-\sin\gamma \to 1$) throughout our study.
The exotic Yukawa couplings for up-type quarks are
\begin{equation}
 \frac{\rho_{ij}}{\sqrt2}\, \bar{u}_{iL}(
  H^0 + {i}\,
  A^0) u_{jR} + {\rm h.c.}, \label{eq:Yuk}
\end{equation}
where, after diagonalizing the mass (and Yukawa) matrices of $h^0$,
$\rho_{ij}$ is in general nondiagonal, but plausibly~\cite{Hou:2017hiw} 
shares the same ``flavor organization'' features of SM,
i.e. trickling down of off-diagonal elements as
reflected in the \emph{observed} quark masses and mixings.
It is by this argument that we expect $\rho_{tt}$ and $\rho_{tc}$
to be ${\cal O}(y_t)$, where $y_t \cong 1$ is the top quark Yukawa coupling in SM.
Likewise, $\rho_{bb}$ and $\rho_{\tau\tau}$ (and $\rho_{\tau\mu}$)
should not be much larger than the respective $y_b$ and $y_\tau$.
Note that $\rho_{ct} \simeq 0$ is demanded by $B$ physics
constraints~\cite{Altunkaynak:2015twa} (see also Ref.~\cite{Chen:2013qta}).

In the following, we set $\rho_{ij} = 0$
except for $\rho_{tt}$ and $\rho_{tc}$.
We plot the $H^0$, $A^0 \to t\bar t$, $t\bar c$ branching ratios
in Fig.~\ref{br-dw} for $\rho_{tt} = 1$ and
 $\rho_{tc} =  0.1$ (solid), 0.5 (dashed) and 1 (dots),
and leading order (LO) total widths 
$\Gamma_{H^0}$ and $\Gamma_{A^0}$ in third panel.
%
We assume $H^0$, $A^0$ and $H^\pm$ to be sufficiently
close in mass, such that decays like $H^0 \to A^0Z^0$, $H^\pm W^\mp$
are kinematically forbidden.
Also not displayed is a small branching into $gg$ final state
close to $t\bar t$ threshold, which does not affect our analysis.
To avoid further model dependence such as
ignoring all other $\rho_{ij}$ and {top} width effects,
we study above $t\bar t$ threshold.

\begin{figure*}[htbp!]
\center
\includegraphics[width=.4 \textwidth]{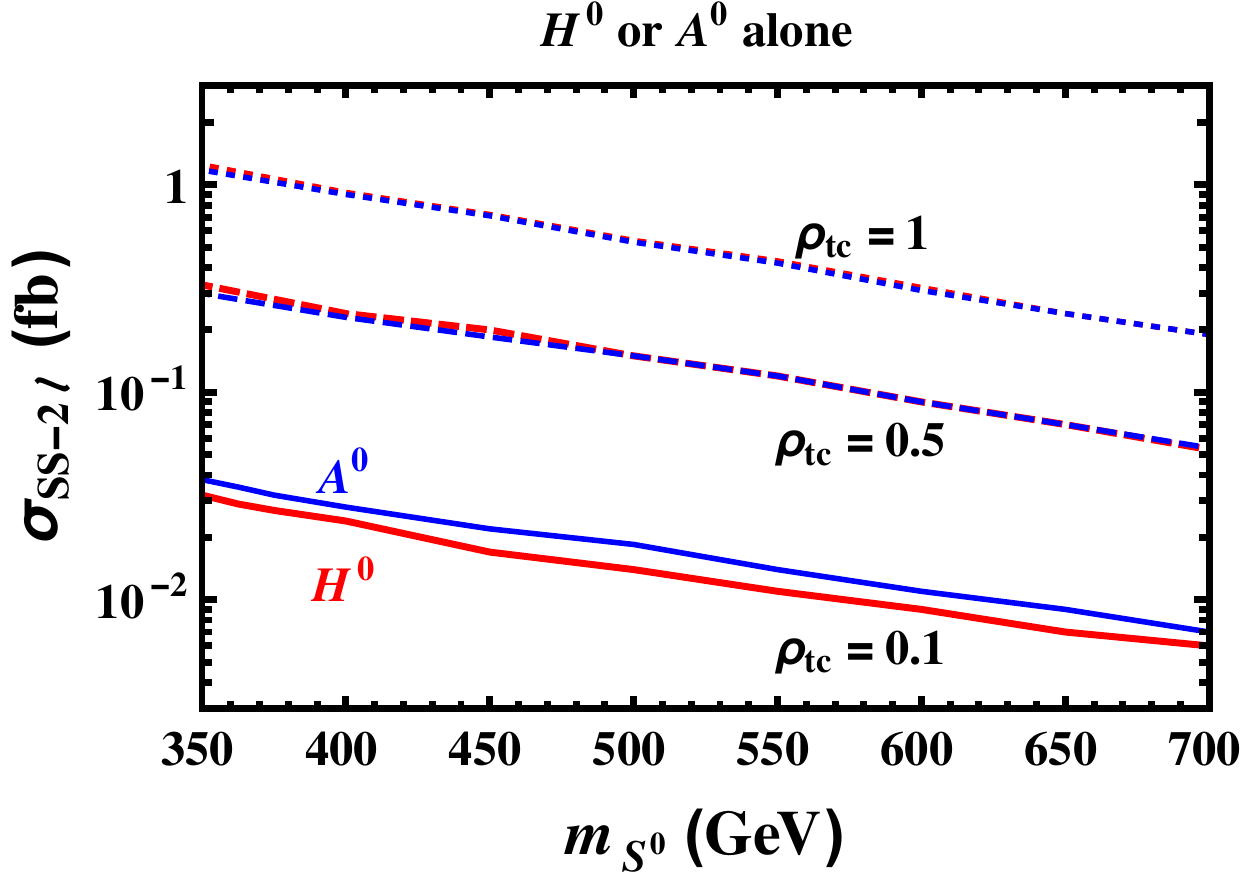}
\includegraphics[width=.39 \textwidth]{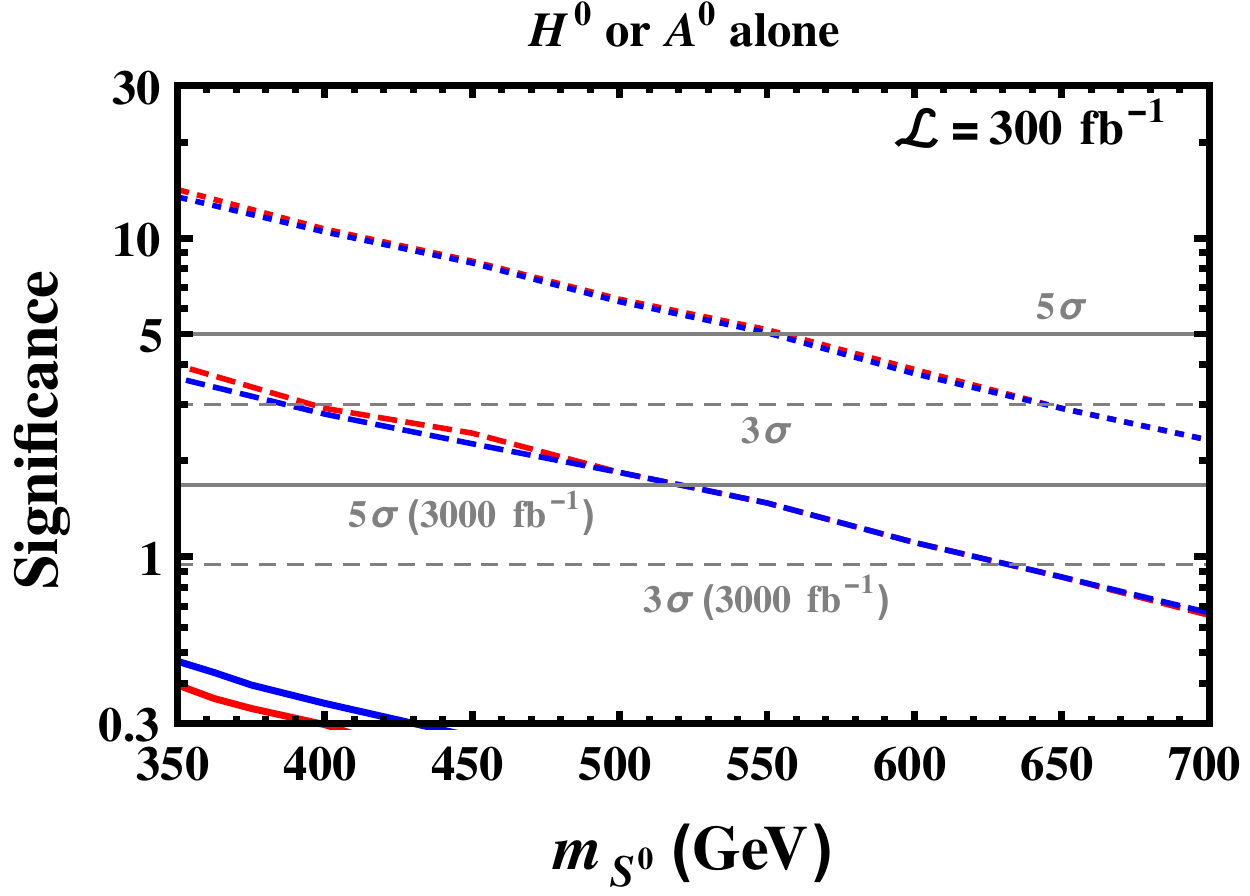}
\caption{
 SS2$\ell$ [left] cross section (fb) and
 [right] significance (for 300 fb$^{-1}$) at $\sqrt s = 14$ TeV. 
 For $\rho_{tc} = 0.1$, $H^0$ curve is lower.
}
\label{ssll}
\end{figure*}

\begin{figure*}[htbp!]
\center
\includegraphics[width=.4 \textwidth]{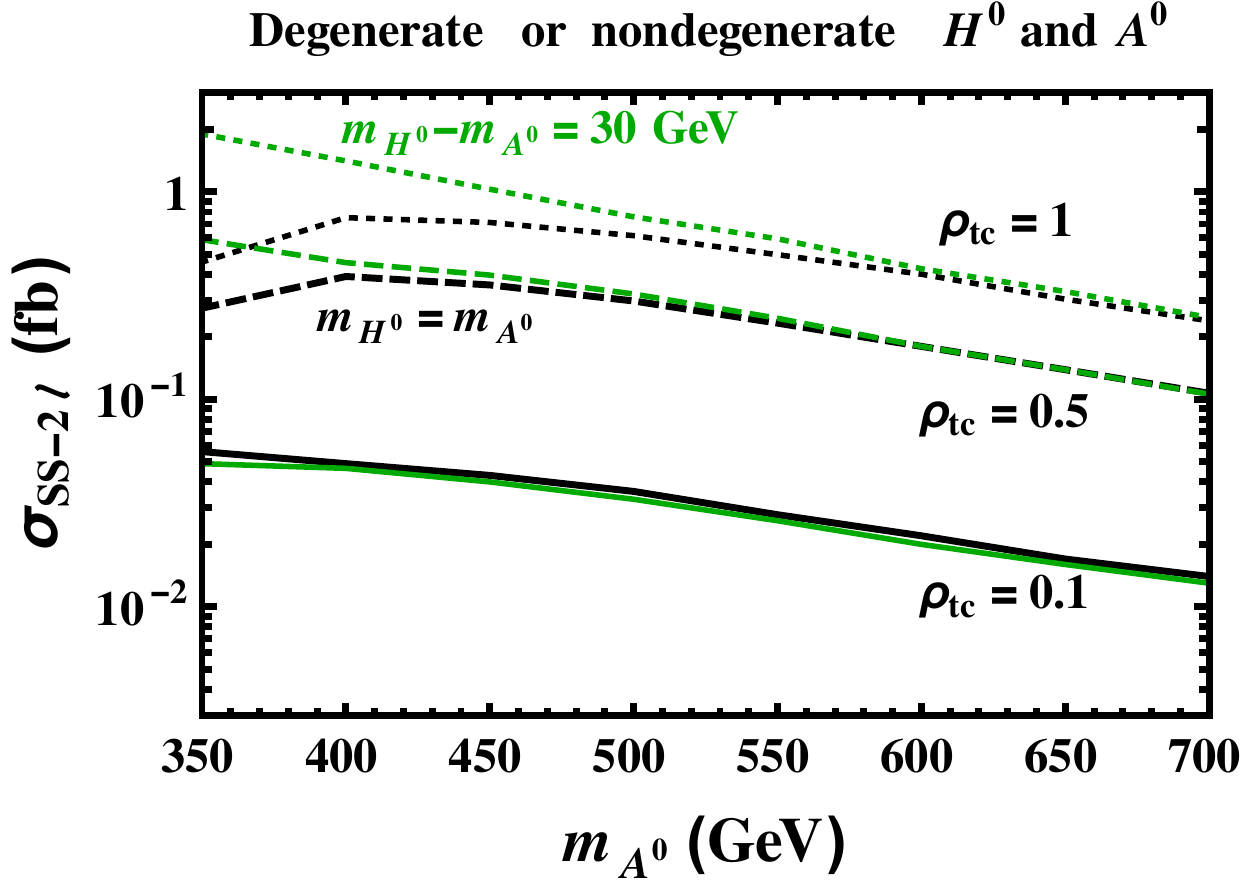}
\includegraphics[width=.39 \textwidth]{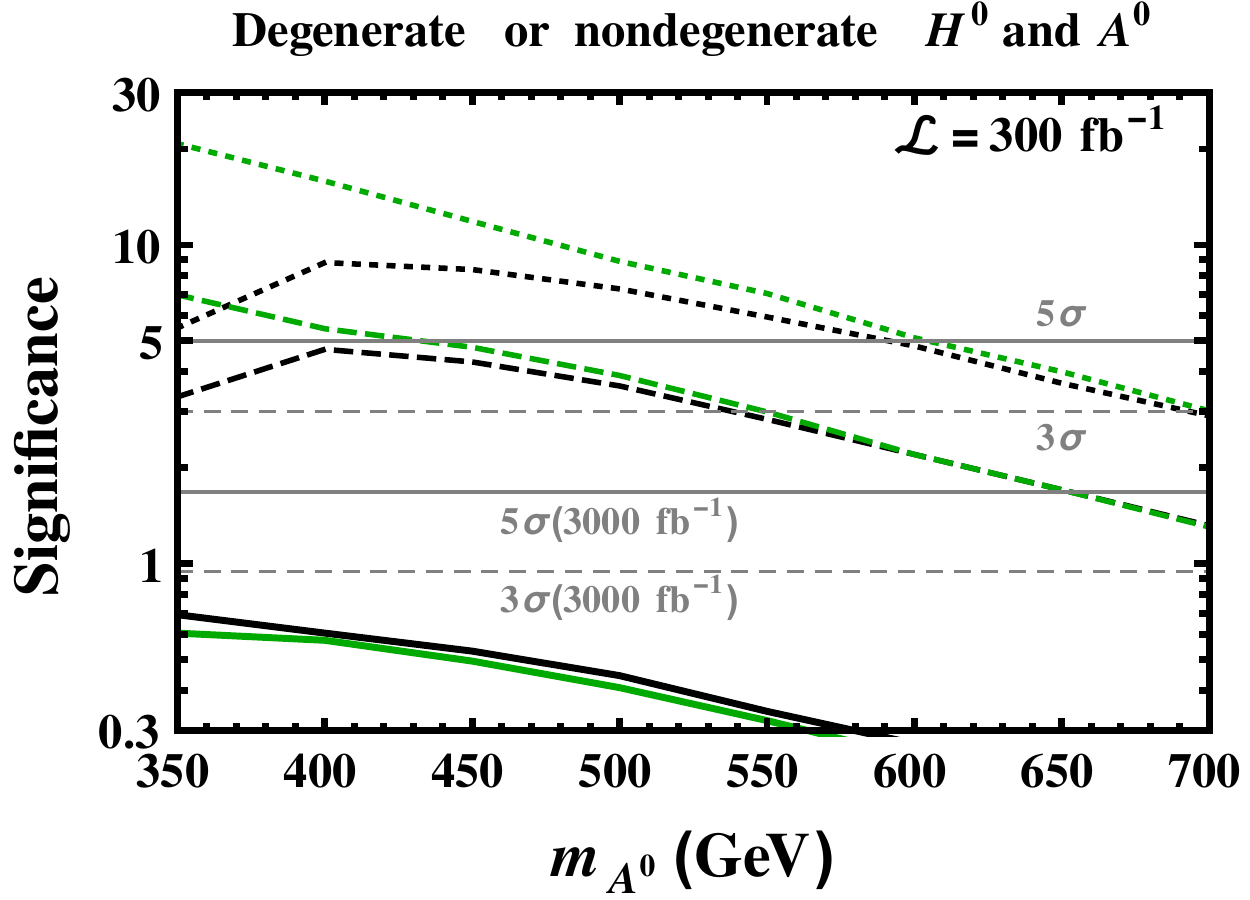}
\caption{
Same as Fig.~\ref{ssll}, but combining $H^0$ and $A^0$ for
$m_{H^0} = m_{A^0}$  (dark black) and 
$m_{H^0} -m_{A^0} =30$ GeV (light green).
}
\label{ssll_degene}
\end{figure*}

\paragraph{Same-sign dilepton.---}

As discussed in Introduction, sizable $\rho_{tt}$ and $\rho_{tc}$
generate $gg \to S^0 \to t\bar t$, $t\bar c$, $\bar tc$, where $S^0 = H^0$, $A^0$.
But the former process is hampered by interference~\cite{Carena:2016npr} with $t\bar t$ background,
while both may suffer in significance~\cite{KFC} because of poor $t\bar t$
or $tj$ mass resolution, where $j$ is a light jet.
Our proposed strategy is to utilize $\rho_{tc} \neq 0$
to search for $cg \to tS^0$, which gives rise to
$tt\bar t$, $tt\bar c$ and $t\bar tc$ plus conjugate final states.
While the true target is triple-top,
an intermediate step would be same-sign dileptons,
which the $tt\bar c$, $\bar t\bar tc$ final states can also feed,
but is rather suppressed in SM.

%
Whether same-sign dileptons or triple-top,
we include $t\bar tS^0$ production~\cite{Craig}
followed by $S^0 \to t\bar t$, $t\bar c$, 
{$\bar t c$.}
Using MadGraph5\_aMC@NLO~\cite{Alwall:2014hca} and
adopting NN23LO1 PDF set~\cite{Ball:2013hta},
the $t\bar tS^0$ cross section is given also in Fig.~\ref{partcross},
where $t\bar tA^0$ is larger than $t\bar tH^0$.
While almost two orders below $tS^0$ for $\rho_{tc} \simeq 1$,
it can feed our signatures,
and would dominate over $tS^0$ for low $\rho_{tc}$~\cite{no_tcS}.
%


The SS2$\ell$ signature is defined as
two leptons with same charge plus at least three jets,
with at least two jets identified as $b$-jets,
and missing transverse energy $E_T^{\rm miss}$.
The SM background processes are $t\bar t Z$, $t\bar t W$,
$tZ +$ jets, $3t + j$, $3t + W$, $4t$, and $t\bar t h$.
The $t\bar t$ and $Z/\gamma^* +$ jets processes, which have
large cross sections, may contribute to background
if the charge of a lepton gets misidentified (charge- or $Q$-flip),
with probability taken as
$2.2\times 10^{-4}$~\cite{ATLAS:2016kjm,Alvarez:2016nrz} in our analysis.

We use MadGraph5\_aMC@NLO to generate event samples at LO 
for $pp$ collisions at $\sqrt{s}=14$ TeV,
interfaced with PYTHIA~6.4~\cite{Sjostrand:2006za} for showering and hadronization,
and adopt MLM matching scheme~\cite{Alwall:2007fs}
for matrix element and parton shower merging.
The event samples 
are fed into 
Delphes~3.4.0~\cite{deFavereau:2013fsa} 
for detector effects.
Eq.~(3) is implemented using FeynRules~2.0~\cite{Alloul:2013bka}.

The selection cuts are as follows.
Transverse momenta ($p_T$) of leading and subleading
leptons $> 25$ GeV and 20 GeV, and
$> 30$ GeV and 20 GeV, respectively for the two {leading} $b$-jets, 
while $E^{\rm miss}_{T} > 30$ GeV.
The pseudo-rapidity of the same-sign leptons and the two leading $b$-jets
should satisfy $|\eta^{\ell}| < 2.5$ and $|\eta^{b}| < 2.5$, respectively.
Separation between  a $b$-jet and a lepton ($\Delta R_{b\ell}$),
any two $b$-jets ($\Delta R_{bb}$), and any two leptons ($\Delta R_{\ell\ell}$)
are required to be $> 0.4$.
We reconstruct jets by anti-$k_T$ algorithm with radius parameter $R=0.6$
and take rejection factors $5$ and $137$ for $c$-jets and light-jets,
respectively~\cite{ATLAS:2014ffa}.
Finally, we require the scalar sum of transverse momenta, $H_T$,
of two leading leptons 
and three leading jets to be $> 300$ GeV.

\begin{table}[b]
\centering
\begin{tabular}{c |c c c c }
\hline\hline
                      Backgrounds                 &  Cross section (fb)      \\
\hline 
                       $t\bar{t}Z$                 & 0.04                \\
                       $t\bar{t}W$                 & 0.72               \\
                       $tZ+$jets                   & 0.001                     \\
                       $3t+j$                      & 0.0002               \\
                       $3t+W$                      & 0.0004               \\
                     {$t\bar t h$}            & {0.024}                      \\
                       $4t$                        & 0.04                  \\
                       $Q$-flip                    & 0.04                     \\
\hline
\hline
\end{tabular}
\caption{Backgrounds for same-sign dilepton at 14 TeV.}
\label{backg_ssll}
\end{table}

\begin{figure*}[htbp!]
\center
\includegraphics[width=.4 \textwidth]{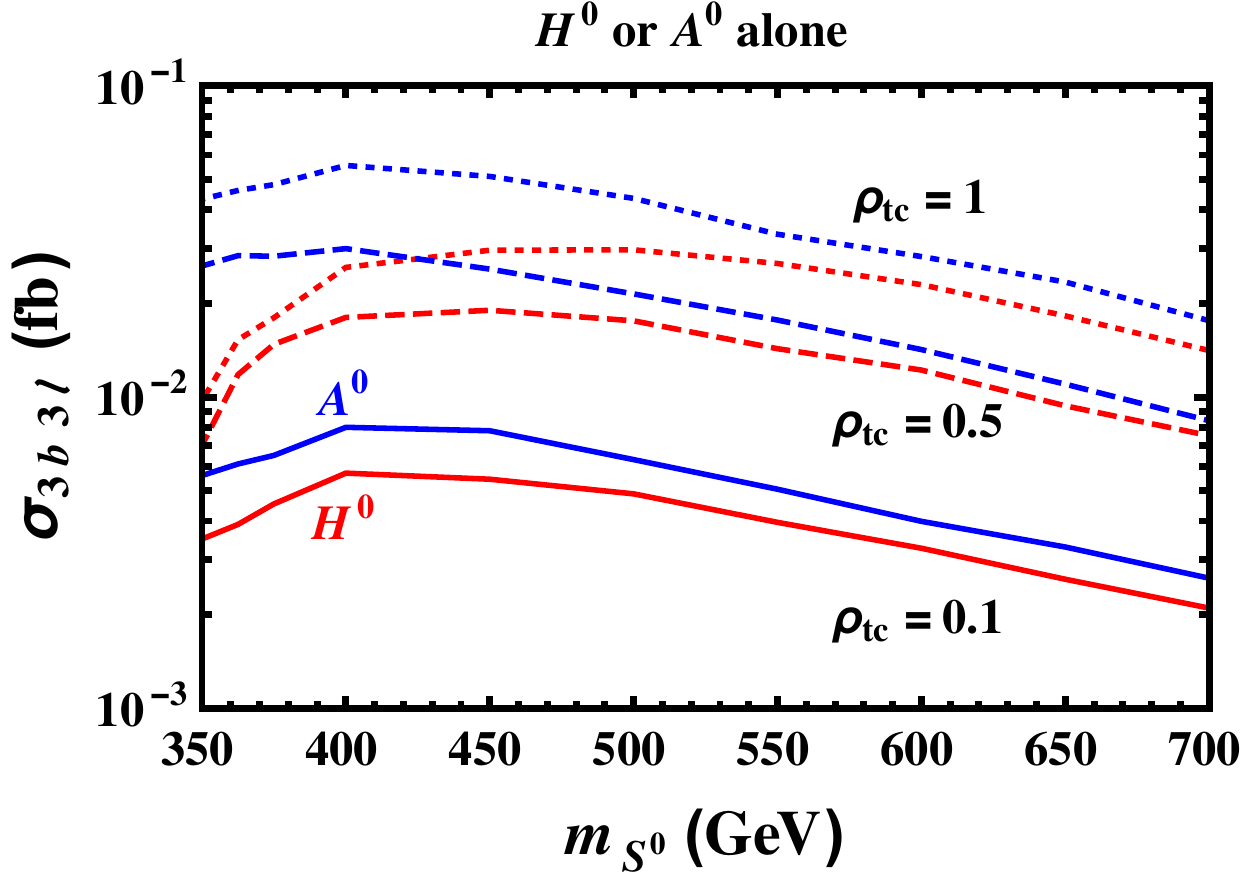}
\includegraphics[width=.39 \textwidth]{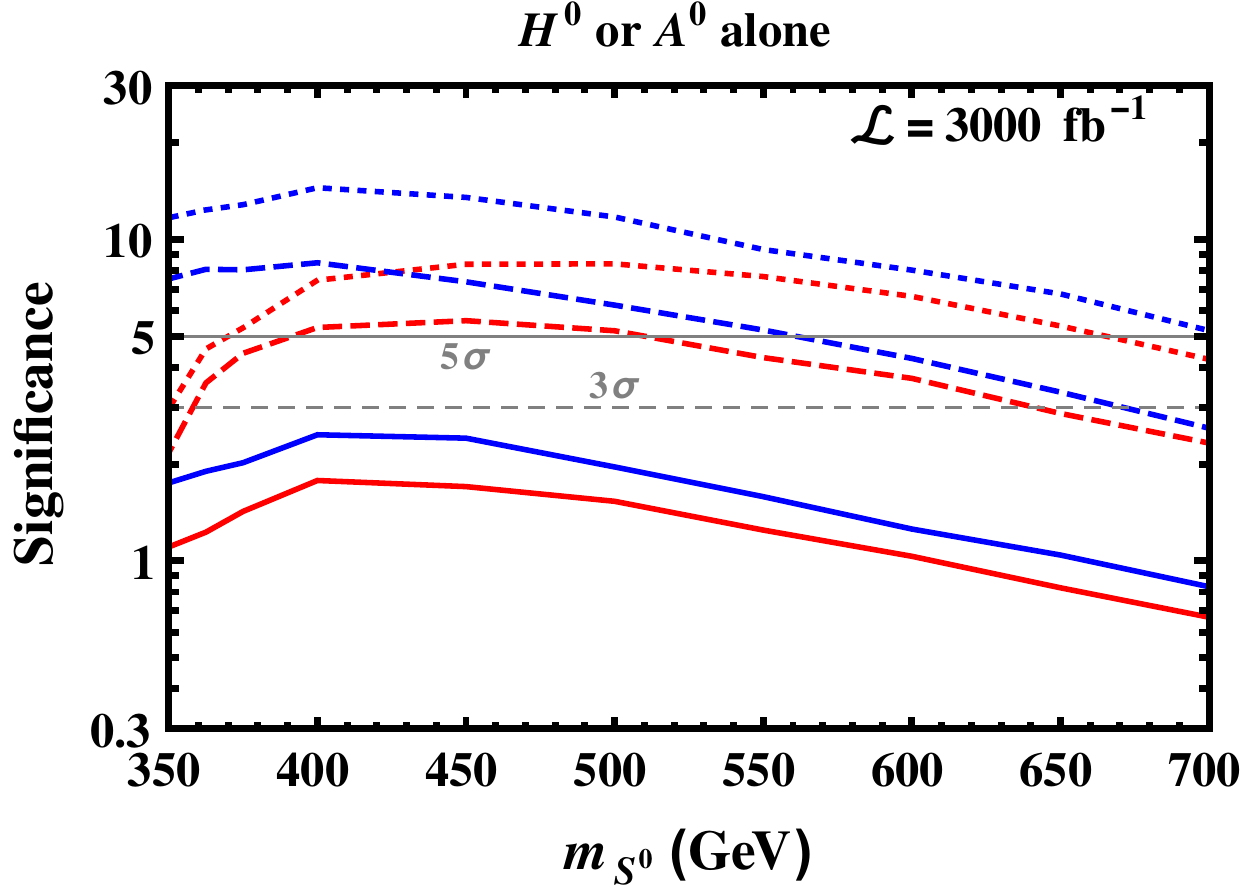}
\caption{
 [Left] Cross sections (fb), and [right] significance (3000 fb$^{-1}$)
 for $3b3\ell$ final state at $\sqrt s = 14$ TeV after selection cuts.
}
\label{3b3l}
\end{figure*}
\begin{figure*}[htbp!]
\center
\includegraphics[width=.4 \textwidth]{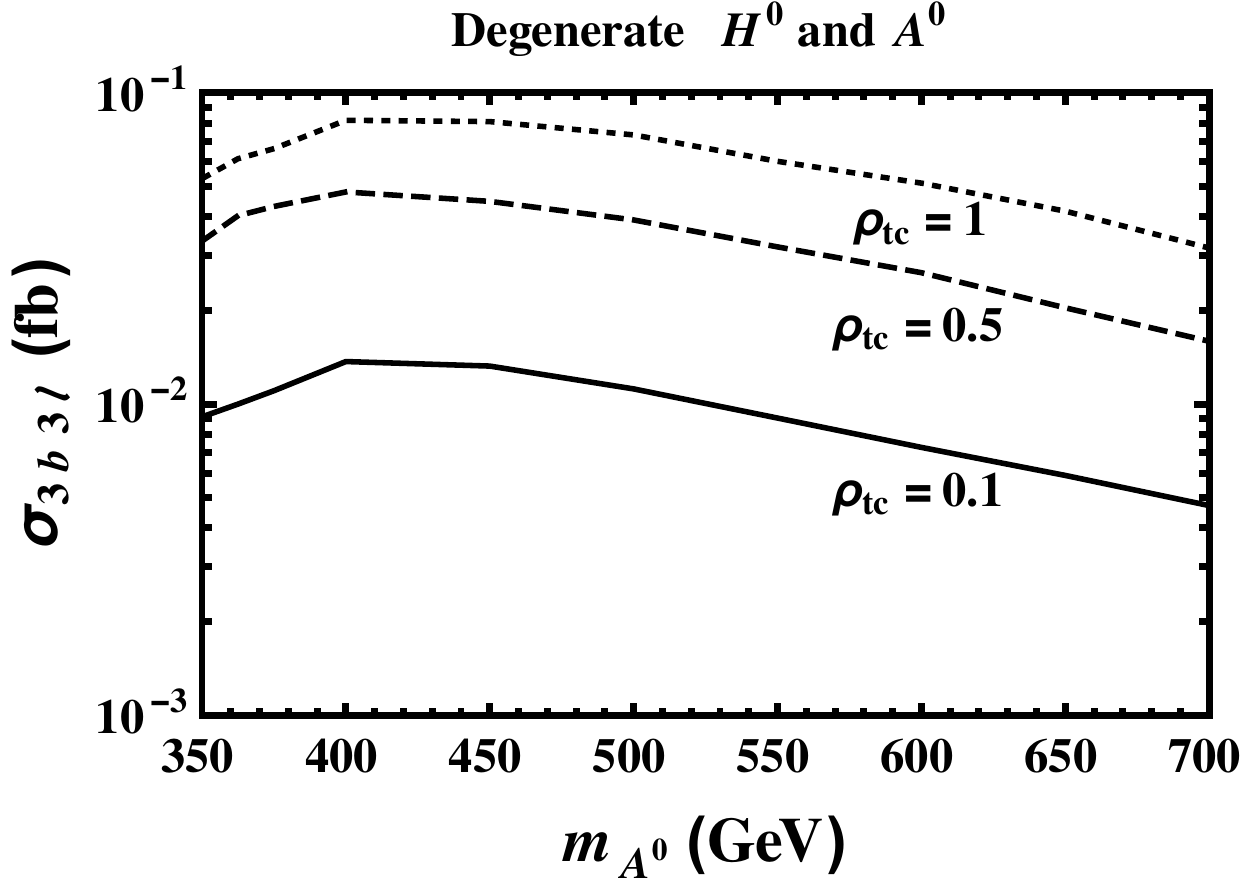}
\includegraphics[width=.39 \textwidth]{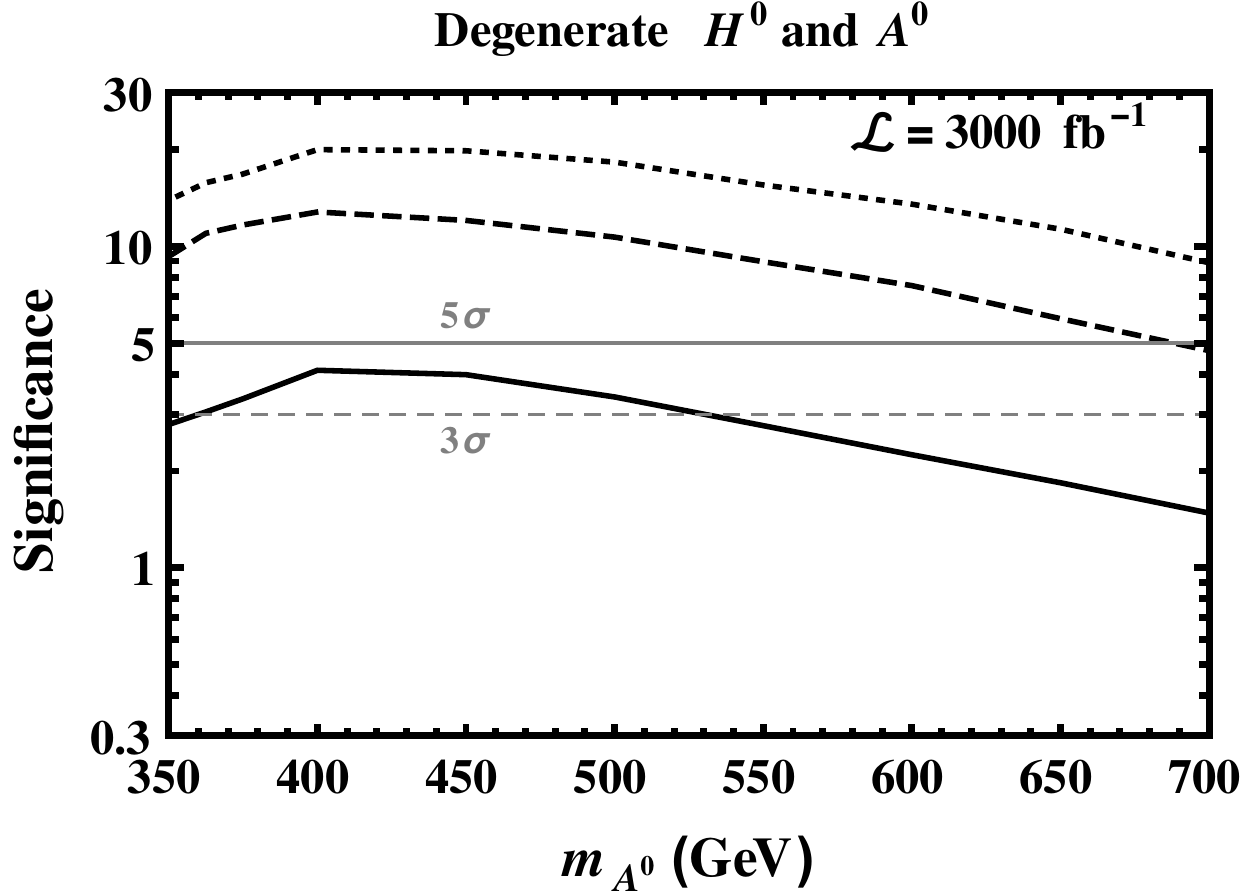}
\caption{Same as Fig.~\ref{3b3l} but combining for $m_{A^0} \approx m_{H^0}$,
where a small mass difference makes little effect.
}
\label{3b3l_degene}
\end{figure*}

The $4t$, $t\bar t Z$,  $tZ +$ jets, $t\bar t W$ and {$t\bar t h$} cross sections at LO are
adjusted to NLO by the factors 2.04~\cite{Alwall:2014hca}, 1.56~\cite{Campbell:2013yla},
1.44~\cite{Alwall:2014hca}, 1.35~\cite{Campbell:2012dh} and 1.27~\cite{twikittbarh}, respectively.
The $t\bar t +$ jets and $Z/\gamma^*$+jets components of $Q$-flip backgrounds are normalized to
NNLO cross sections by $1.84$~\cite{twiki} and $1.27$ respectively, where
the latter is obtained using FEWZ 3.1~\cite{Li:2012wna}.
The correction factors for conjugate processes are assumed to be the same for simplicity.

The signal cross sections after selection cuts for $\rho_{tt} = 1$
and $\rho_{tc} = 0.1$, 0.5 and 1 are plotted in Fig.~\ref{ssll}[left],
with backgrounds given in Table~\ref{backg_ssll}.
For $\rho_{tc} = 1$ and 0.5, the $tS^0$ process predominates,
with both $S^0 \to t\bar c$ and $t\bar t$ contributing,
and it is hard to distinguish between the curves for $H^0$ and $A^0$ 
in the logarithmic scale except near $t\bar t$ threshold, where 
the $H^0$ curve exceeds the $A^0$ one up to 10\% for $\rho_{tc}=0.5$.
For $\rho_{tc} = 0.1$, the $tS^0$ process drops below $t\bar tS^0$,
with $t\bar tA^0$ slightly higher than $t\bar tH^0$ (Fig.~1),
so $t\bar tt\bar t$ dominates the cross section
except near $t\bar t$ threshold.

There are also ``nonprompt'' backgrounds which may be significant.
The CMS study of SS$2\ell$ signature~\cite{Sirunyan:2017uyt},
{with slightly different cuts},
finds nonprompt background at $\sim1.5$ times the $t\bar{t}W$ background.
These backgrounds are not properly modeled in our Monte Carlo simulations.
We simply add a nonprompt background that is 1.5 times 
$t\bar t W$ after selection cuts to the overall background.

Using $\mathcal{Z} = \sqrt{2[ (S+B)\ln( 1+S/B )-S ]}$~\cite{Cowan:2010js}, 
we estimate the signal significance with 300 fb$^{-1}$
and plot in Fig.~\ref{ssll}[right], 
with 5$\sigma$ discovery (or 3$\sigma$ evidence) potential shown as 
horizontal gray lines, including scaling to 3000 fb$^{-1}$ by $\sqrt{\mathcal{L}}$.
For $\rho_{tc} = 0.1$,
there is practically no significance at 300 fb$^{-1}$,
because the signal is small compared with even $t\bar tW$ background.
Since both $H^0$ and $A^0$ are produced, 
$H^0$ and $A^0$ interference effects are important.
The two cases of degenerate $H^0$ and $A^0$, 
and small mass splitting $m_{H^0}-m_{A^0}=30$ GeV, 
are illustrated in Fig.~\ref{ssll_degene},
with discussion deferred until later.

\paragraph{Triple-top.---}
Our triple-top signature, denoted as $3b3\ell$, is defined as
at least three leptons 
and at least three jets,
of which at least three are $b$-jets, and $E^{\rm miss}_T$.

Dominant SM backgrounds are $t\bar t Z+$jets and $4t$,
with $t\bar t Wb$, $tZjb$, $3t +j$, $3t + W$, and $t\bar t h$ subdominant.
The $t\bar t +$jets process can contribute if a
jet gets misidentified as a lepton, with probability taken as
$\epsilon_{\rm fake} = 10^{-4}$~\cite{Aad:2016tuk, Alvarez:2016nrz}.
We do not include nonprompt backgrounds
as they are not properly modeled in Monte Carlo simulations.
Unlike the SS2$\ell$ signature,
the three hard leptons plus high $b$-jet multiplicity, along with
$Z$-pole veto and $H_T$ cut may reduce such contributions significantly.
The $4t$, $t\bar t Z +$ jets,  $tZjb$, $t\bar t Wb$ and {$t\bar th$} cross sections
at LO are adjusted to NLO by same factors as in previous section,
i.e. 2.04, 1.56, 1.44, 1.35 and 1.27 respectively, and likewise $1.84$
for $t\bar t +$ jets. Conjugate processes are assumed to have same correction factors.

The selection cuts are as follows.
Three leading leptons and three leading $b$-jets each satisfy
$p^{\ell}_{T} > 25$ GeV and $p^{b}_T > 20$ GeV, respectively.
Pseudo-rapidity and separation requirements as well as $E_T^{\rm miss}$
are as in SS2$\ell$ process.
The scalar sum of transverse momenta $H_T$ of
{three leading leptons and three leading $b$-jets} should be $> 320$ GeV.
To reduce $t\bar t Z +$ jets background,
we veto~\cite{CMS:2017uib} $76~\mbox{GeV}< m_{\ell\ell} < 95~\mbox{GeV}$
for same flavor, opposite charged lepton pair, and if more than one such pair, 
the veto is applied to the one closest to $m_Z$.

The signal cross sections after selection cuts are plotted 
as before in Fig.~\ref{3b3l}[left],
with backgrounds listed in Table~\ref{backg_3l3b}.
For $\rho_{tc} = 1$ and 0.5, the $A^0$ cross section is higher than $H^0$
because ${\cal B}(A^0 \to t\bar t) > {\cal B}(H^0 \to t\bar t)$ (Fig.~\ref{br-dw}), 
which also explains the slower turn on for $H^0$ as $m_H$ is raised.
For $\rho_{tc} = 0.1$, the $t\bar tS^0$ process dominates,
with higher cross section for $A^0$ (Fig.~1).
The significance, estimated now for 3000~fb$^{-1}$, is plotted in Fig.~\ref{3b3l}[right].
We give the incoherent sum of $H^0$ and $A^0$ results 
for degenerate case in Fig.~\ref{3b3l_degene}, 
which is verified by parton level calculation with $H^0$--$A^0$ interference 
by MadGraph for $pp \to tH^0/A^0 \to tt\bar t$ and 
$pp \to t\bar t H^0/A^0 \to t\bar t t\bar t$, $t\bar t t \bar c$, $t\bar t \bar t c$.
Thus, a small mass splitting makes little effect.

\begin{table}[t!]
\centering
\begin{tabular}{c |c c c c }
\hline\hline
                      Backgrounds                 &  Cross section (fb)      \\
\hline
                       $t\bar t Z+$jets            & 0.0205~~(0.0026)               \\
                       $t\bar{t}Wb$                & 0.0017~~(0.0015)               \\
                       $tZjb$                      & \hspace{-.7cm}0.0002~~($-$)                      \\
                       $3t+j$                      & 0.0001~~(0.0001)               \\
                       $3t+W$                      & 0.0004~~(0.0003)               \\
                      {$t\bar t h$ }           & {0.0015~~(0.0013) }                      \\
                       $4t$                        & 0.0232~~(0.0209)                    \\
                       $t\bar t$+jets (fake)       & 0.0026~~(0.0025)                     \\
\hline
\hline
\end{tabular}
\caption{
Backgrounds for $3b 3\ell$ at 14 TeV,
where numbers in parentheses are with additional $Z$-pole veto.}
\label{backg_3l3b}
\end{table}

%
\paragraph{Discussion and Conclusion.---}

Let us understand our results.
In Fig.~\ref{ssll}, $cg \to tS^0$ dominates for $\rho_{tc} = 1$ and 0.5,
and there is little distinction between $H^0$ or $A^0$,
and likewise for $tt\bar t$ or $tt\bar c$ in giving rise to same-sign dilepton.
Hence the $H^0$ and $A^0$ curves are almost identical.
The high significance for $\rho_{tc} \sim 1$
suggests it should already be relevant in LHC Run 2.
There is, however, some subtlety.
In Fig.~\ref{ssll_degene}, the combined result for degenerate $H^0$ and $A^0$ 
are shown as dark (black) lines. 
Here, $cg \to tH^0 \to tt\bar c$ and $cg \to tA^0 \to tt\bar c$ amplitudes cancel each other, 
up to ${H^0}$ and ${A^0}$ width difference (see Fig.~\ref{br-dw}[right]).
This can be understood from Eq.~(\ref{eq:Yuk}), where the 
$A^0$ amplitude gains a factor of $i^2=-1$ to the  $H^0$ one~\cite{tri-top-intf}.
This cancellation depends crucially on our assumption of $|\rho_{tc}| \gg |\rho_{ct}|$,
and leads to the drop near $t\bar t$ threshold for $\rho_{tc}=1,\, 0.5$.
For small mass splitting such as $m_{H^0} -m_{A^0} = 30$ GeV 
(light (green) lines in Fig.~\ref{ssll_degene}), the cancellation
is relaxed for smaller masses, but is still effective for higher masses due to larger width.
It is therefore $tt\bar t$ that is the dominant contribution,
even with some splitting for higher masses.

CMS has searched~\cite{Sirunyan:2017uyt} for SS2$\ell$ type of events 
using 35.9 fb$^{-1}$ data at 13 TeV, which can already constrain $\rho_{tc}$. 
In the CMS analysis, signal regions (SRs) are defined according to lepton $p_T$, 
number of $b$-jets, 
$E_T^{\rm miss}$, $H_T$~\cite{CMS_HT}, etc.,
then aggregated into a smaller number of exclusive SRs.
One of these, ExSR5, which resembles our SS2$\ell$ discussion,
is defined as follows: 
two same-sign leptons in the high-high (HH) category (each $p_T^\ell> 25$ GeV), 
at least two jets with two $b$-tagged (each with $p^b_T > 25$ GeV), 
$ E_T^{\rm miss} \in [50, 300]$ GeV and $H_T < 1125$ GeV,
with the additional requirement of $m_T^{\rm min} < 120$ GeV (smallest transverse
mass constructed from a lepton and $\vec{p}_T^{\rm \,miss}$) 
for events with $H_T > 300$ GeV.
The observed 152 events is consistent with SM expectation of $137 \pm 25$.

\begin{figure}[t]
\center
\includegraphics[width=.39 \textwidth]{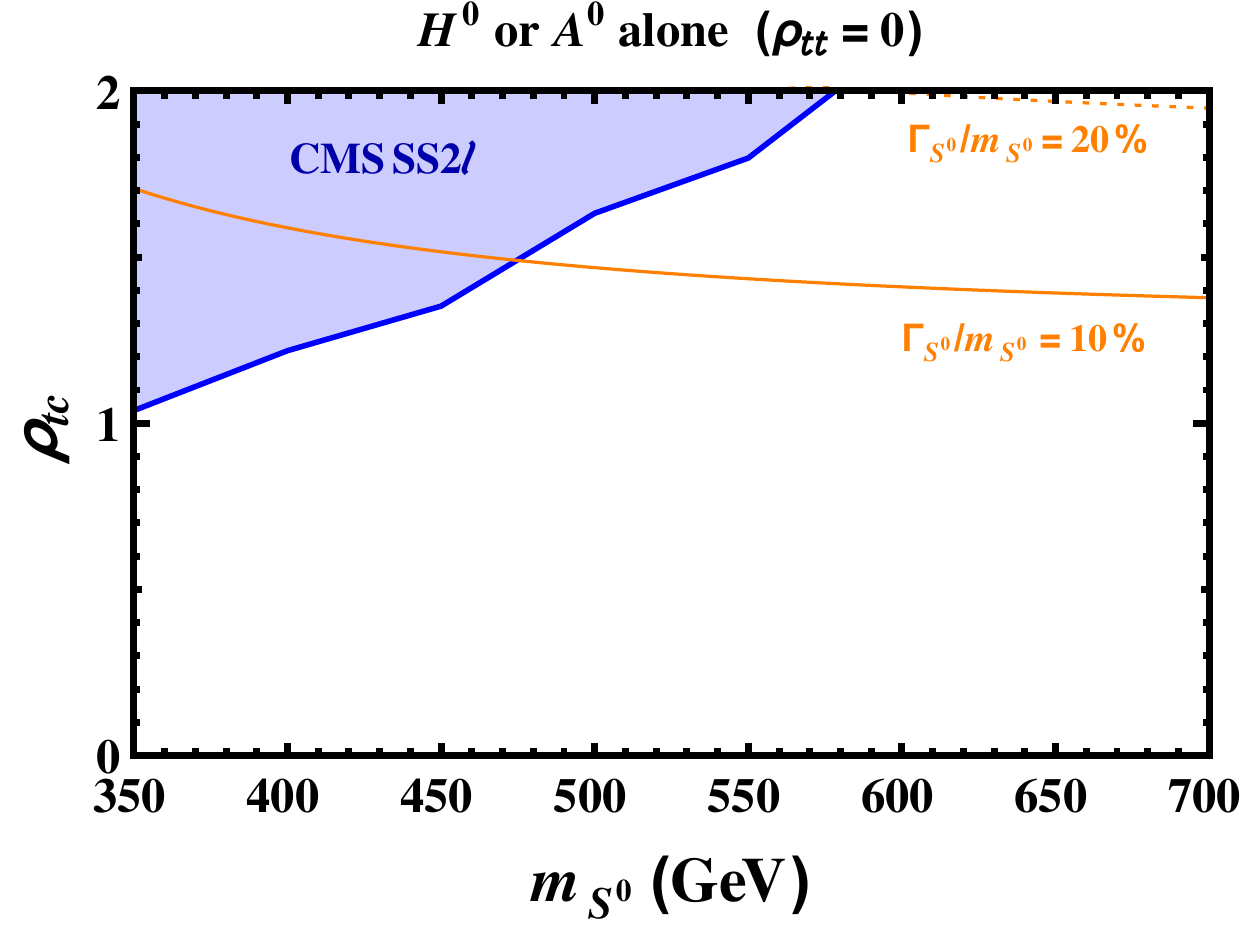}
\caption{
Region of 2$\sigma$ exclusion for $\rho_{tc}$ vs $m_{S^0}$ with $\rho_{tt}=0$
by CMS~\cite{Sirunyan:2017uyt} same-sign dilepton search with 35.9~fb$^{-1}$ at 13 TeV,
in ``ExSR5'' exclusive signal region with two $b$-jets (see text).
The result is the same for $H^0$ or $A^0$ alone.
}
\label{rhotc_const}
\end{figure}

{
Following the CMS cuts, we estimate the contribution 
from $pp \to t H^0 \to tt\bar c$ (and conjugate) for $\rho_{tt} = 0$ and $\rho_{tc} = 1$, 
then scale by $|\rho_{tc}|^2$ assuming narrow $H^0$ width.
Demanding the sum with SM expectation 
to not exceed the observed number of events within $2\sigma$, 
the obtained upper limit is shown in Fig.~\ref{rhotc_const}.
We do not display above $\rho_{tc}= 2$, where the width gets rather large
(orange curves in Fig.~\ref{rhotc_const} for $\Gamma_{H^0}/m_{H^0} = 0.1, 0.2$).
The same limits apply to the case with $A^0$ alone.
If $H^0$ and $A^0$ are degenerate with $\rho_{tt} = 0$,
their effect in the ExSR5 vanishes.  
For small mass splitting 
the BSM contribution is nonzero, but the simple scaling of
cross section by $| \rho_{tc} |^2$ cannot be applied 
for the $H^0$-$A^0$ interference term.
We note that in any case 
the ExSR5 is not optimized for our SS2$\ell$ search,
as it sums over SRs allowing less energetic jet activities, 
bringing in an SR with rather large number of expected ($70\pm 12$) vs 
observed ($90$) events.
One could optimize the exclusive SRs for better limit on $\rho_{tc}$,
but it is better left to experiment.

With Run 2 data not yet probing our target parameter range, 
we return to our SS2$\ell$ analysis.
Since one would pick up both $H^0$ and $A^0$ effects, 
we see from Fig.~\ref{ssll_degene} that $5\sigma$ discovery reach 
at 300 fb$^{-1}$ extends to 600 GeV 
for $\rho_{tc} = 1$.
For $\rho_{tc} = 0.5$, 
$5\sigma$ discovery would reach 430 GeV only,
and only if there is some splitting between  $H^0$ and $A^0$:
{in this low mass range, 
$tH^0$ and $tA^0$ cancellation} 
would weaken the significance when closer to degenerate.
Below $\rho_{tc} = 0.5$, the situation quickly deteriorates,
but the large 3000 fb$^{-1}$ data at HL-LHC can extend $5\sigma$ discovery
beyond 600 GeV, and looks promising.
Thus, same-sign dileptons can tell us
whether the $\rho_{tc} \sim 1$ mechanism~\cite{Fuyuto:2017ewj} for EWBG is allowed.

%
We would advocate, however,  that ``triple-top'' search is more informative.
Here, $S^0\to t\bar t$ decay is needed for $cg \to tS^0$ to contribute,
hence is less sensitive just above $t\bar t$ threshold.
As cross sections are smaller due to more exquisite selection cuts,
we give results for 3000 fb$^{-1}$.
We see from Fig.~\ref{3b3l_degene}[right] that $5\sigma$ discovery reach 
extends to 700 GeV and even higher for $\rho_{tc} \gtrsim 0.5$. 
The range shrinks as $\rho_{tc}$ drops, but   
even for $\rho_{tc}\sim 0.1$, one could get some hint (at $3\sigma$ or higher)
up to $m_{H^0} \cong m_{A^0} \sim$ 500 GeV. 
%
This can be compared with SS2$\ell$ case,
which effectively has no sensitivity for $\rho_{tc} \sim 0.1$ (Fig.~\ref{ssll_degene}[right]).

The above assumes $\rho_{tt} \sim 1$.
A smaller $\rho_{tt}$ would reduce the triple-top signal,
but also reduce the SS2$\ell$ signal at higher mass
due to $tH^0$--$tA^0$ cancellation. 
Since $tt\bar t$ dominates both SS$2\ell$ and $3b3\ell$ in high mass range, 
one can check consistency between the two signatures if mass splitting is small.
For lower masses with a small mass splitting, 
the relative strength of triple-top and same-sign top  at high luminosity,
could allow one to extract information on relative strength of $\rho_{tc}$ vs $\rho_{tt}$, 
assuming discovery.
This would help us understand whether EWBG is more 
driven by $\rho_{tt}$ or $\rho_{tc}$.
Although we have elucidated the effect of exotic $H^0$ and $A^0$ scalars,
mass reconstruction would not be easy,
and further study would be needed, especially for HL-LHC.
Can the $gg \to H^0/A^0 \to t\bar c,\, t\bar t$ processes bear fruit eventually,
especially for mass reconstruction?

We have assumed nearly degenerate heavy scalars, which need not be the case.
Finite splittings could lead to $H^0 \to A^0Z^0$, $H^\pm W^\mp$ (or reverse)
decays, which would dilute our signatures but enrich the program.
One charm of new $\rho_{tt}$ and $\rho_{tc}$ Yukawa couplings 
is their intrinsic complexity, which is why they can drive EWBG~\cite{Fuyuto:2017ewj}.
More studies are needed to probe these CPV phases.
Our proposal is thus only a first step of a large program.

In conclusion, motivated by electroweak baryogenesis and alignment,
we suggest searching for $cg \to tH^0$, $tA^0$
that arise from a new Yukawa coupling $\rho_{tc}$
in 2HDM without discrete $Z_2$ symmetry.
If $\rho_{tc} \sim 1$ hence could {drive} EWBG,
we could discover same-sign dileptons with 300 fb$^{-1}$.
Given that $\rho_{tt}$ is favored as driver for EWBG,
triple-top search at HL-LHC could cover full mass range
up to 700 GeV for $\rho_{tt} \sim 1$,
but $\rho_{tc}$ needs to be not much smaller than 0.5.
Advancing from same-sign dileptons to triple-top at the LHC,
we may test our understanding of matter{-antimatter} asymmetry of the Universe.


\vskip0.2cm
\noindent{\bf Acknowledgments} \
We thank K.-F. Chen for discussions,
and the support of grants NTU-ERP-106R8811 and NTU-ERP-106R104022,
and MOST 105-2112-M-002-018, 106-2811-M-002-187 and 106-2112-M-002-015-MY3.
MK thanks S. Iguro and K. Tobe for correspondence.
WSH thanks G.N. Taylor for hospitality.


\end{document}